\documentclass{PoS}
\usepackage{amsmath}

\title{An effective theory for QCD with an axial chemical potential}

\ShortTitle{An effective theory for QCD with axial chemical potential}

\author{\speaker{Xumeu Planells}\\
Departament d'Estructura i Constituents de la Mat\`eria and Institut de Ci\`encies del Cosmos (ICCUB), Universitat de Barcelona, Spain\\
E-mail: \email{xumeu@icc.ub.edu}}

\author{Alexander A. Andrianov and Vladimir A. Andrianov\\
V. A. Fock Department of Theoretical Physics, Saint-Petersburg State University, Russia\\
E-mail: \email{andrianov@icc.ub.edu}, \email{v.andriano@rambler.ru}}
\author{Domenec Espriu\\
Departament d'Estructura i Constituents de la Mat\`eria and Institut de Ci\`encies del Cosmos (ICCUB), Universitat de Barcelona, Spain\\
E-mail: \email{espriu@ecm.ub.es}}

\abstract{We consider the low energy realization of QCD in terms of meson fields when an axial chemical potential is present; a situation that may be relevant in heavy ion collisions. We shall demonstrate that the presence of an axial charge constitutes an explicit source of parity breaking. The eigenstates of strong interactions do not have a definite parity and interactions that would otherwise be forbidden compete with the familiar ones. In this work, we first focus on scalars and pseudoscalars that are described by a generalized linear sigma model; and next, we give some hints on how the Vector Meson Dominance model describes the vector sector.}

\FullConference{The XXI International Workshop High Energy Physics and Quantum Field Theory\\
June 23 - June 30, 2013 \\
Saint Petersburg Area, Russia}

\begin{document}

\section{Motivation of local parity breaking}

It is well known that parity is a well established global symmetry of strong interactions. However, there are reasons to believe that it may be broken in a finite volume since the Vafa-Witten theorem \cite{VW} does not apply for $\mu\neq 0$. Accordingly, some time ago, it was proposed that the QCD vacuum can possess metastable domains leading to $P$ violation.

\medskip

Two different types of works have been of special interest in the study of the origin of this possible parity breaking effect. On the one hand, there is an important list of contributions that were devoted to the investigation of the existence of stable $P$- and $CP$-odd "pion" condensates (see \cite{pioncond}). On the other hand, local large topological fluctuations also had a relevant role in a hot environment \cite{topfluc}. This work deals with the last of these hypothesis.

\begin{figure}[h!]
\begin{minipage}[t]{6.9cm}
\vspace{-12.2em}\qquad It is conjectured that the presence of a non-trivial angular momentum (or magnetic field) in peripheral heavy ion collisions (HIC) leads to the so-called Chiral Magnetic Effect (CME) \cite{kharz}, where a separation of electrical charges takes place thus implying a clear signature for local parity breaking (LPB). However, in central collisions, fluctuations in the topological charge lead to an ephemeral phase with axial chemical potential $\mu_5 \neq 0$. In Fig. \ref{QCDphases} we show where the LPB phase may exist within the known QCD phase diagram.
\end{minipage}\qquad \begin{minipage}[t]{7.45cm}
\centering
\includegraphics[scale=0.3]{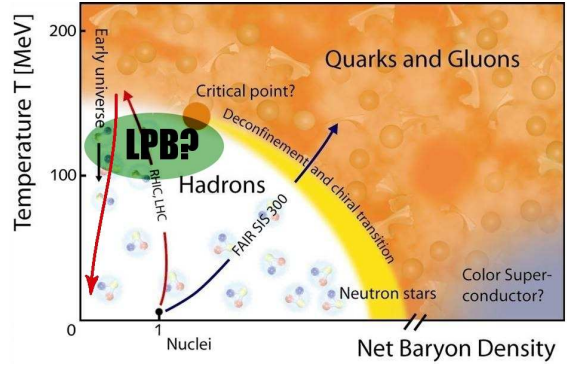}
 \vspace{-2em}
\caption{QCD phase diagram depending on the temperature and density with the hypothetical region where LPB may occur.\label{QCDphases}}
\end{minipage}
\end{figure}

\section{Axial baryon charge and axial chemical potential}

The strong interaction shows a highly non-trivial vacuum that allows different topological sectors, separated by high energy barriers. These non-equivalent configurations may be connected due to large quantum fluctuations of the vacuum state in the presence of a hot medium via sphaleron transitions. A topological charge $T_5$ may arise in the fireball
\[T_5=\frac{1}{8\pi^2}\int_{\mbox{\footnotesize vol.}}d^3x\varepsilon_{jkl}\mbox{Tr}\left (G^j\partial^k G^l-i\frac23G^jG^kG^l\right )\]
and survive for a sizeable lifetime. During this period, one can control the jump in the topological charge $\langle \Delta T_5 \rangle$, a gauge invariant object, by introducing into the QCD Lagrangian a topological chemical potential $\mu_\theta$ through $\Delta \mathcal L_{top} =\mu_\theta \Delta T_5$, where
\[\Delta T_5 = T_5(t_f) - T_5(0) = \frac{1}{8\pi^2}\int^{t_f}_0 dt\int_{\mbox{\footnotesize vol.}}d^3x \mbox{Tr}\left (G^{\mu\nu} \widetilde G_{\mu\nu}\right ).\]
In the chiral limit, the integration of the local PCAC, broken by the gluon anomaly, predicts a conserved axial charge $Q^q_5$ during $\tau_{\mbox{\footnotesize fireball}}\sim 5\div 10$ fm connected to $T_5$ via
\[\frac d{dt}\left (Q_5^q-2N_f T_5\right )\simeq 0, \quad Q_5^q=\int_{\mbox{\footnotesize vol.}}d^3x\bar q\gamma_0\gamma_5q = \langle N_L - N_R\rangle.\]
Thus, the presence of a non-trivial topological charge in the finite volume is directly related to the emergence of a quark axial charge, which has a characteristic oscillation time governed by the inverse quark masses. Therefore, for $u$, $d$ quarks  $1/m_q \sim 1/5$ MeV$^{-1}$ $\sim 40$ fm, so the oscillation can be neglected and during the fireball lifetime we may consider the chiral charge to be approximately constant. However, for the $s$ quark one has $1/ m_s \sim 1/150$ MeV$^{-1}$ $\sim 1$ fm $\ll \tau_{\mbox{\footnotesize fireball}}$ and even if a topological charge persists during fireball lifetime, the mean value of strange quark chiral charge is close to zero due to essential left-right oscillations.

\medskip

As a result, for $u$ and $d$ quarks, QCD with a background topological charge leads to the generation of an axial chemical potential $\mu_5$, conjugate to the quark axial charge $Q_5^q$, which can be related to the topological charge and chemical potential in the following way:
\begin{equation}
\langle \Delta T_5 \rangle \simeq \frac{1}{2N_f} \langle Q_5^q \rangle \, \Longleftrightarrow \, \mu_5 \simeq \frac{1}{2N_f} \mu_\theta, \qquad \Delta {\cal L}_{top}= \mu_\theta\Delta T_5 \, \Longleftrightarrow \, \Delta {\cal L}_q = \mu_5 Q_5^q\nonumber.
\end{equation}

\section{Effective scalar/pseudosalar meson theory with $\mu_5$}

This section is devoted to show how the parity breaking effect is introduced in hadron physics and its main results. If we focus in the scalar meson sector, the spurion technique can be used by taking $\mu_5$ as the time component of some external axial-vector field. In this work, we will consider $\mu_5$ to be an isosinglet. Therefore the derivative operator is extended in the following way
\begin{equation}
D_\nu \Longrightarrow  D_\nu - i \{{\bf I}_q\mu_5 \delta_{0\nu}, \cdot \}=D_\nu - 2i{\bf I}_q\mu_5 \delta_{0\nu}.
\nonumber
\end{equation}
Note the breaking of Lorentz symmetry when the time component of the external field is selected. In a medium where parity is broken, two new processes are likely to appear inside the fireball: the decays $\eta,\eta^\prime \to \pi\pi$ that are strictly forbidden in QCD on parity grounds. And even more, they might reach thermal equilibrium. In order to investigate such an effect, we construct a generalized $\Sigma$ model with the following effective Lagrangian:
\begin{align}\label{Lagr}
\mathcal L=&\frac 14 \mbox{Tr}\left (D_\mu HD^\mu H^\dag\right )+\frac b2 \mbox{Tr}\left [M(H+H^\dag)\right ]
+\frac{M^2}2 \mbox{Tr}\left (H H^\dag\right ) - \frac{\lambda_1}2 \mbox{Tr}\left [(HH^\dag)^2\right ]-\frac{\lambda_2}4 \left [\mbox{Tr}\left (H H^\dag\right )\right ]^2\nonumber\\
&+\frac c2 (\mbox{det}H + \mbox{det}H^\dag)+ \frac{d_1}2 \mbox{Tr}\left [M(H H^\dag H+H^\dag H H^\dag)\right ]+
\frac{d_2}2 \mbox{Tr}\left [M(H+H^\dag)\right ] \mbox{Tr}\left (H H^\dag\right )
\end{align}
where the main building blocks are
\begin{equation}
H=\xi \Sigma\xi,\quad \xi=\exp\left(i\frac{\Phi}{2f}\right ),\quad \Phi=\lambda^a\phi^a,\quad \Sigma=\lambda^b\sigma^b.
\nonumber
\end{equation}
The v.e.v. of the neutral scalars are defined as $v_i=\langle\Sigma_{ii}\rangle$ where $i=u,d,s$, and satisfy the following gap equations:
\[M^2v_i-2\lambda_1 v_i^3-\lambda_2\vec v^2v_i+c\frac{v_uv_dv_s}{v_i}=0.\]

For further purposes we need the non-strange meson sector (composed of pions, $\eta_q$, $\sigma(600)$, $\vec a_0(980)$) and $\eta_s$. The relation between $\eta_q$, $\eta_s$ and the genuine degrees of freedom $\eta$ and $\eta'$ is given by a plain rotation of the fields. In summary, we have
\begin{equation}
\Phi=\!\!\!\begin{pmatrix}
\eta_q+\pi^0 & \sqrt{2}\pi^+ & 0\\
\sqrt 2 \pi^- & \eta_q-\pi^0 & 0\\
0 & 0 & \sqrt 2\eta_s
\end{pmatrix}\!\!\!, ~\Sigma=\!\!\!\begin{pmatrix}
v_u+\sigma+a_0^0 & \sqrt 2 a_0^+ & 0\\
\sqrt 2 a_0^- & v_d+\sigma-a_0^0 & 0\\
0 & 0 & v_s
\end{pmatrix}\!\!\!,
\begin{pmatrix}
\eta_q\\
\eta_s
\end{pmatrix}\!\!=\!\!\begin{pmatrix}
\cos\psi & \sin\psi\\
-\sin\psi & \cos\psi
\end{pmatrix}\!\!\!\!\begin{pmatrix}
\eta\\
\eta'
\end{pmatrix}\!\!.\nonumber
\end{equation}

For $\mu_5=0$, we assume $v_u=v_d=v_s=v_0\equiv f_\pi\approx 92$ MeV. 
The coupling constants (in units of MeV) are fitted to phenomenology assuming isospin symmetry via a $\chi^2$ minimization with the subroutine \texttt{MINUIT}:
\begin{gather}
b=-3510100/m,~~ M^2=1255600,~~ c=1252.2,~~ \lambda_1=67.007, \nonumber\\
\lambda_2= 9.3126,~~ d_1=-1051.7/m,~~ d_2=523.21/m, 
\nonumber
\end{gather}
where $m\equiv m_q=(m_u+m_d)/2$ and $m/m_s\simeq 1/25$ is used.

\medskip

Next, we present a simple case of mixing due to LPB in the isotriplet sector with $\pi$ and $a_0$. The kinetic and mixing terms in the Lagrangian are given by
\[\mathcal L=\frac 12 (\partial a_0)^2+\frac 12 (\partial \pi)^2-\frac12 m_1^2 a_0^2-\frac12 m_2^2 \pi^2-4\mu_5 a_0 \dot{\pi},\]
where
\begin{align}
\nonumber m_1^2=&-2[M^2-2(3\lambda_1+\lambda_2)v_q^2-\lambda_2 v_s^2-cv_s+2(3d_1+2d_2)mv_q+2d_2m_sv_s+2\mu_5^2],\\
\nonumber m_2^2=&\frac{2m}{v_q}\left [b+(d_1+2d_2)v_q^2+d_2v_s^2\right ].
\end{align}

\begin{figure}[h!]
\centering
 \includegraphics[scale=0.28]{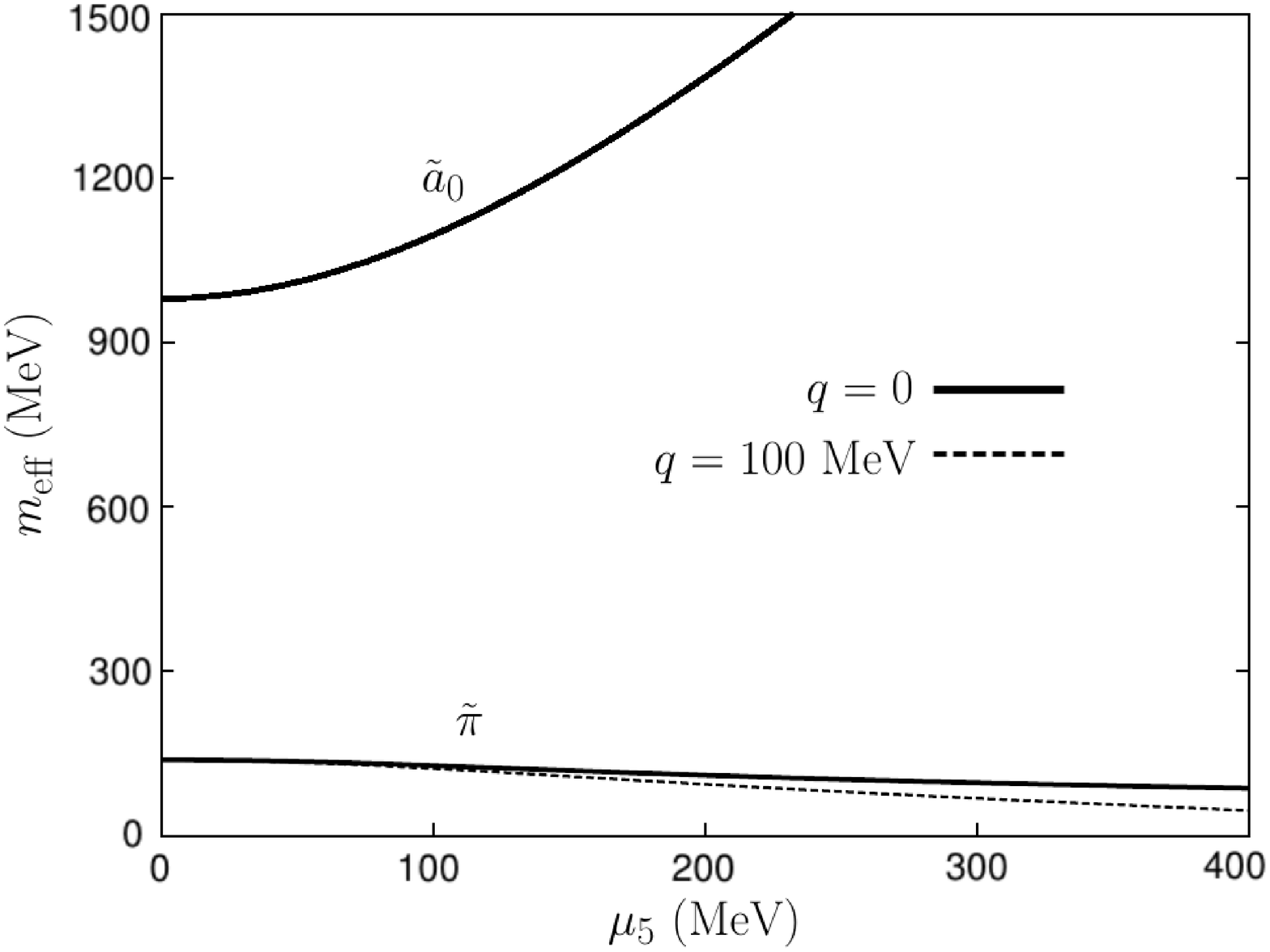}\quad\includegraphics[scale=0.28]{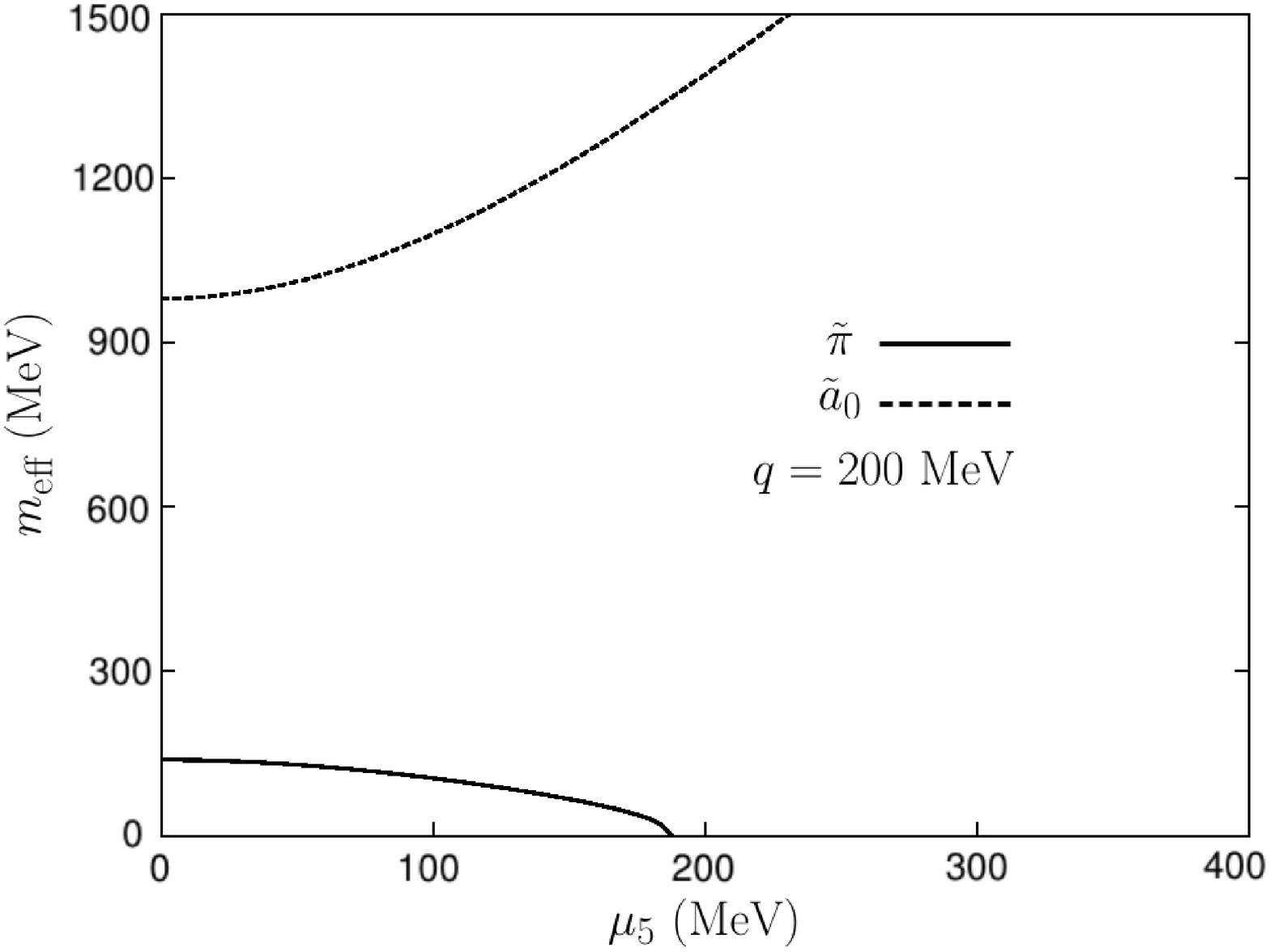}
  \vspace{-0.5em}
\caption{Effective mass dependence on $\mu_5$ for $\tilde\pi$ and $\tilde a_0$. Left panel: comparison of masses at rest and at low momentum $q=100$ MeV. Right panel: same masses at $q=200$ MeV, where the $\tilde\pi$ mass becomes tachyonic.}\label{mass_api}
\end{figure}

After diagonalization in the momentum representation, the new (momentum-dependent) eigenstates are defined as $\tilde\pi$ and $\tilde a_0$. Note that both states reduce to the known ones in the limit $\mu_5=0$, say $\tilde\pi(\mu_5=0)=\pi$ and equivalently for $a_0$. In the left panel of Fig. \ref{mass_api} we see how the $\tilde\pi$ mass decreases as a function of $\mu_5$ for a particle at rest and for $q=100$ MeV while $\tilde a_0$ mass shows an enhancement, but $\mu_5$ has to be understood as a perturbatively small parameter.

In the right panel of Fig. \ref{mass_api}, it can be shown that for high energies (or momenta) bigger than a certain threshold $k^c_{\tilde\pi}\equiv m_1m_2/(4\mu_5)$, the in-medium $\tilde \pi$ goes tachyonic. Nevertheless, energies are checked to be always positive so no vacuum instabilities arise. No appreciable differences occur for the $\tilde{a}_0$ mass.

\medskip

We now turn to the isosinglet sector, where the mixing of $\eta$, $\sigma$ and $\eta'$ is studied. The kinetic and mixing terms in the Lagrangian are given by
\begin{align}
\nonumber \mathcal L=&\frac 12[(\partial\sigma)^2+(\partial \eta_q)^2+(\partial \eta_s)^2]-\frac 12m_3^2 \sigma^2-\frac 12m_4^2\eta_q^2-\frac 12m_5^2\eta_s^2-4\mu_5\sigma\dot{\eta}_q-2 \sqrt{2}cv_q\eta_q\eta_s,
\end{align}
where
\begin{gather}
\nonumber m_3^2=-2(M^2-6(\lambda_1+\lambda_2)v_q^2-\lambda_2 v_s^2+cv_s+6(d_1+2d_2)mv_q+2d_2m_sv_s +2\mu_5^2),\\
\nonumber m_4^2=\frac {2m}{v_q}\left [b+(d_1+2d_2)v_q^2+d_2v_s^2\right ]+2cv_s, \quad m_5^2=\frac{2m_s}{v_s}[b+2d_2v_q^2+(d_1+d_2)v_s^2]+\frac{cv_q^2}{v_s}.
\end{gather}

\begin{figure}[h!]
\centering
 \includegraphics[scale=0.28]{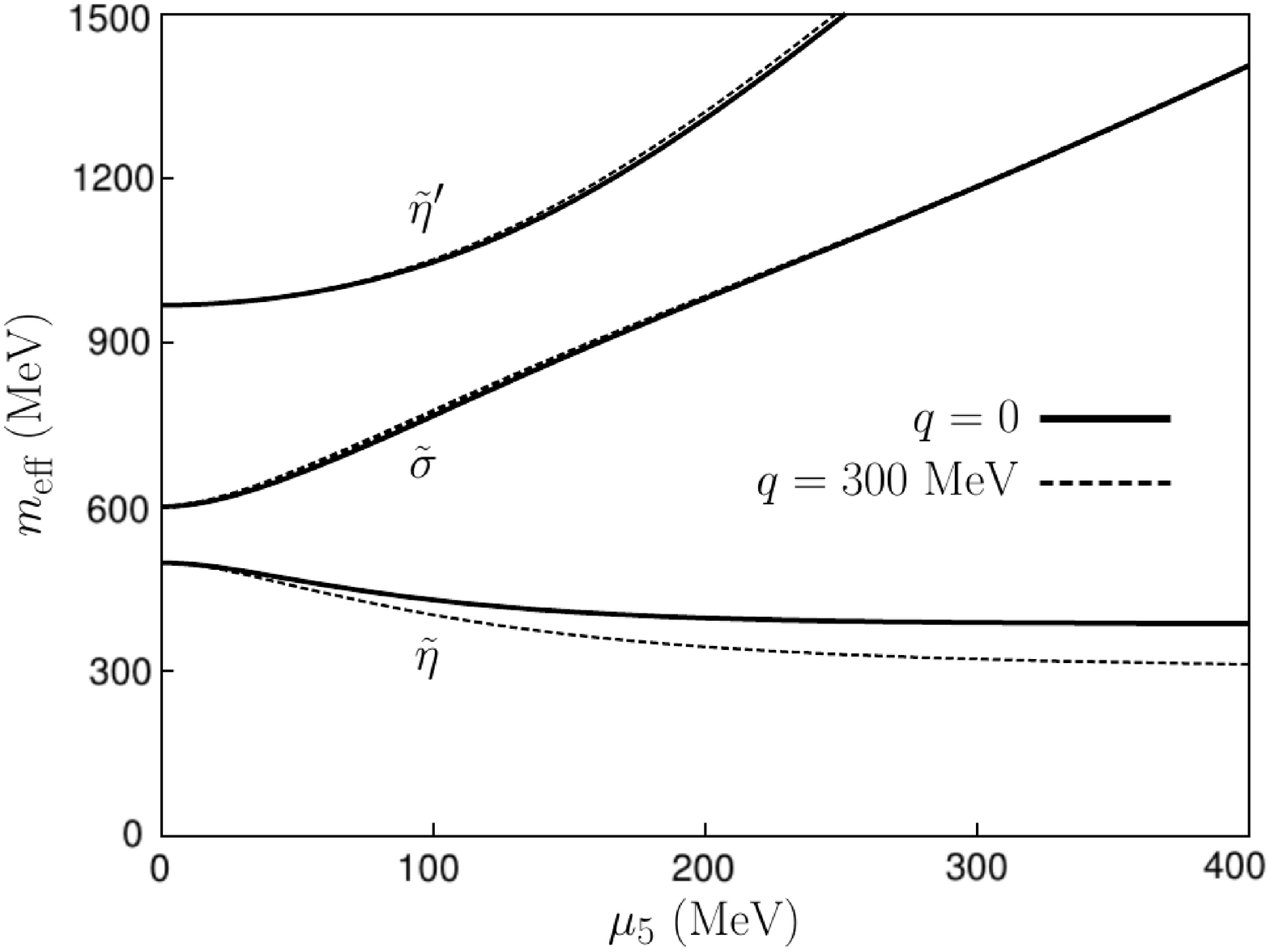}\quad\includegraphics[scale=0.28]{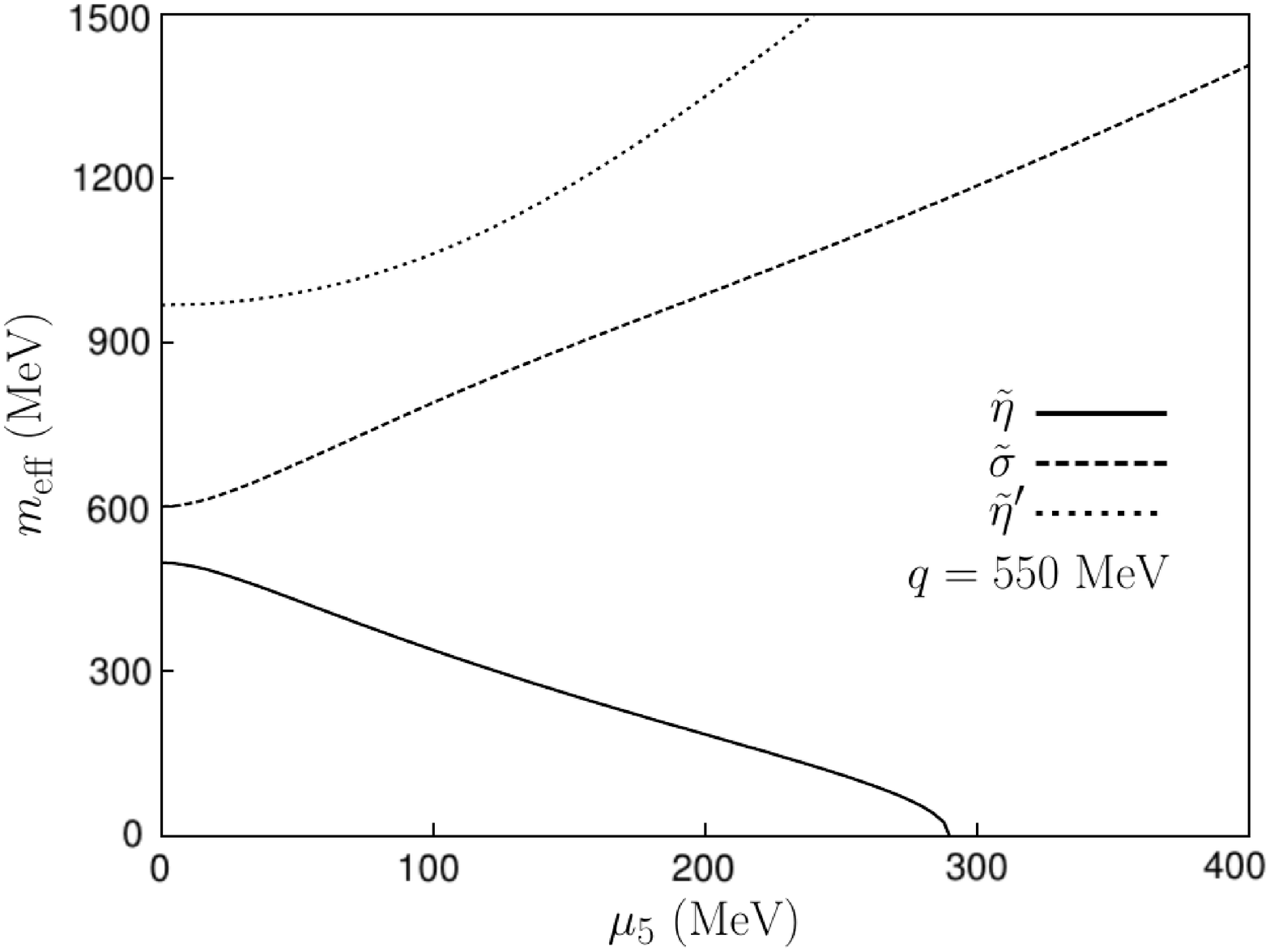}
  \vspace{-0.5em}
\caption{Effective mass dependence on $\mu_5$ for $\tilde\eta$, $\tilde\sigma$ and $\tilde\eta'$. Left panel: comparison of masses at rest and at low momentum $q=300$ MeV. Right panel: masses at $q=550$ MeV, where the $\tilde\eta$ mass goes tachyonic.}\label{mass_singl}
\end{figure}

After diagonalization, and following the notation introduced before, the new eigenstates are defined as $\tilde\sigma$, $\tilde\eta$ and $\tilde\eta^\prime$. In the left panel of Fig. \ref{mass_singl} we find a similar behaviour as the one seen in the triplet sector for the particles at rest and at $q=300$ MeV. Now $\tilde\eta$ decreases with respect to $\mu_5$ while $\tilde\sigma$ and $\tilde\eta'$ increase. Boosting more these eigenstates, we find in the right panel of Fig. \ref{mass_singl} that again for high enough energies (bigger than  the threshold $k^c_{\tilde\eta}\equiv \frac{m_3}{4\mu_5m_5}\sqrt{m_4^2m_5^2-8c^2v_q^2}$), the in-medium $\tilde\eta$ goes tachyonic.

The cubic couplings from our Lagrangian in Eq. \eqref{Lagr} can be extracted and used to calculate the widths $\tilde\eta,\tilde\sigma,\tilde\eta'\to\tilde\pi\tilde\pi$. The important pieces are the following ones:
\begin{gather}
\nonumber \mathcal L_{\sigma aa}=2[(3d_1+2d_2)m-2(3\lambda_1+\lambda_2)v_q]\sigma \vec a_0^2, \quad \mathcal L_{\sigma a\pi}=-\frac{4\mu_5}{v_q}\sigma\vec a_0\dot{\vec{\pi}},\\
\nonumber \mathcal L_{\sigma\pi\pi}=\frac 1{v_q^2}\left [(\partial\vec\pi)^2v_q-(b+3(d_1+2d_2)v_q^2+d_2v_s^2)m\vec\pi^2\right ]\sigma,  \quad \mathcal L_{\eta aa}=-\frac{2 \mu_5}{v_q}\dot \eta_q \vec a_0^2,\\
\nonumber \mathcal L_{\eta a\pi}=\frac 2{v_q^2}\vec a_0[\partial \eta_q \partial \vec\pi v_q-(b+
(3d_1+2d_2)v_q^2+d_2v_s^2)m\eta_q\vec\pi], \quad  \mathcal L_{\eta\pi\pi} = 0.
\end{gather}
After the diagonalization just explained in both the isosinglet and triplet sectors, one replaces the initial degrees of freedom $\{\eta_q,\eta_s,\sigma\}$ and $\{\pi, a_0\}$ to the new ones $\{\tilde\eta,\tilde\sigma,\tilde\eta'\}$ and $\{\tilde\pi,\tilde a_0\}$. Hence, the widths may be numerically computed.

\medskip

\begin{figure}[h!]
\centering
%\vspace{-0.5em}
 \includegraphics[scale=0.28]{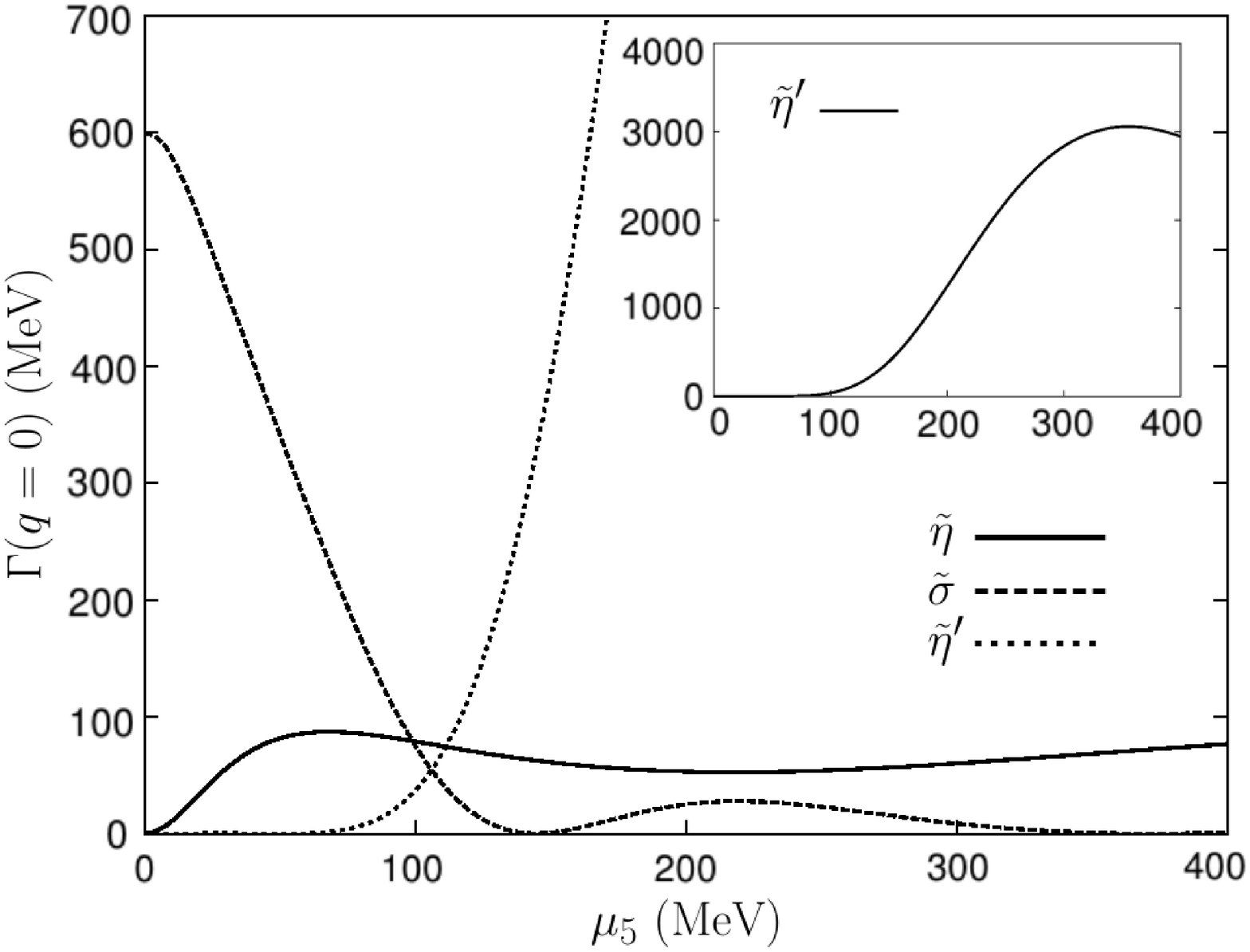}\quad \includegraphics[scale=0.28]{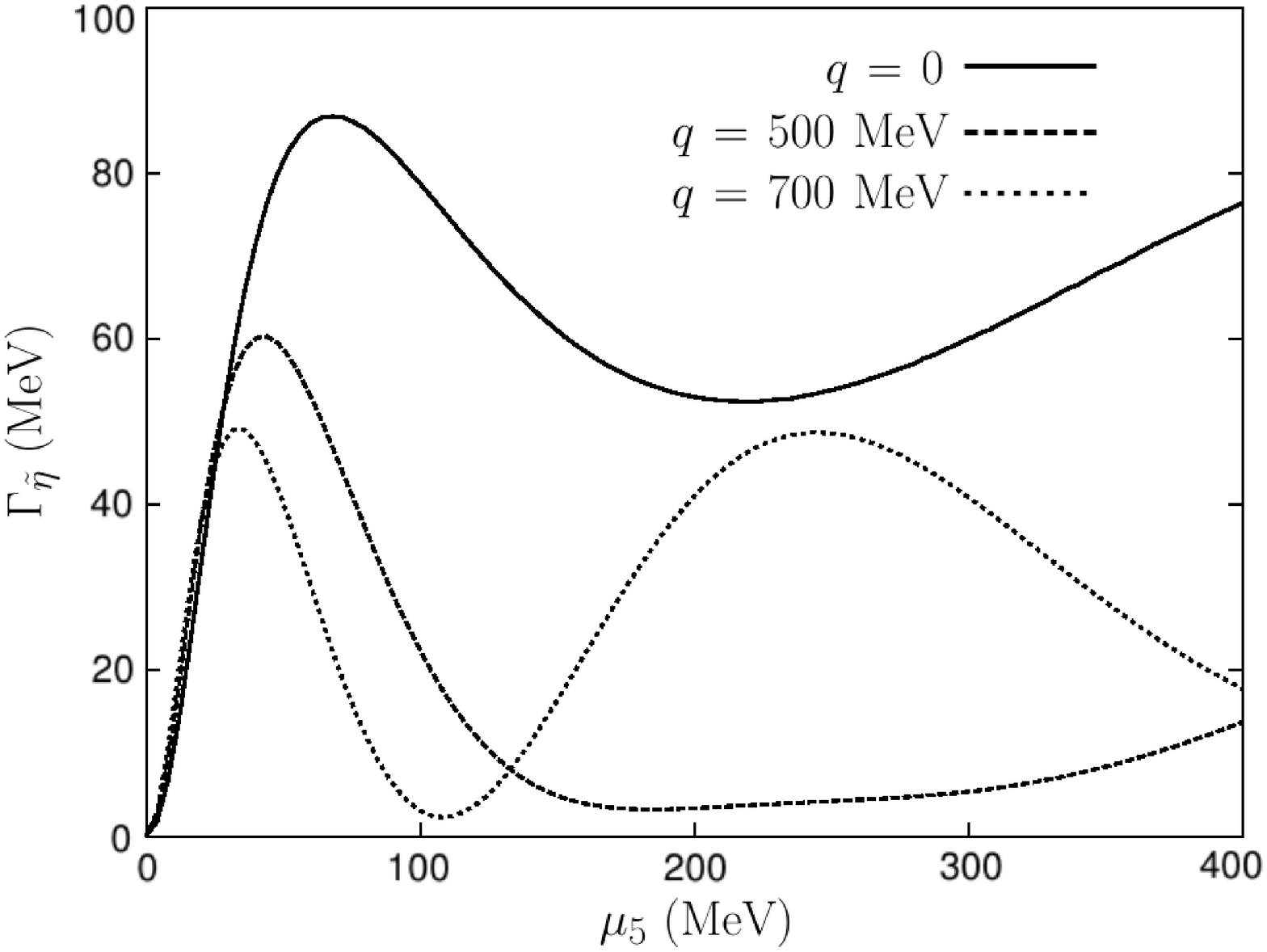}
 \vspace{-0.5em}
\caption{Left panel: $\tilde\eta$, $\tilde\sigma$ and $\tilde\eta'$ widths at rest depending on $\mu_5$. Right panel: $\tilde\eta$ width depending on $\mu_5$ for different values of the incoming 3-momentum: $q=0,500,700$ MeV.}\label{gamma_eta}
\end{figure}

First of all, in the left panel of Fig. \ref{gamma_eta} we present these results at the rest frame of the decaying particle. Down to $\mu_5=50$ MeV, $\tilde\eta$ exhibits a smooth behaviour with an approximate average width of around 60 MeV, what corresponds to a mean free path of 3 fm, which is smaller than the typical fireball size of $L_{\text{fireball}}\sim 5\div 10$ fm. As a result, a regeneration of such particles into (distorted) pions may take place in a HIC, as it is well known to happen with the $\rho^0$ meson, and they could possibly thermalize within the pion gas.
On the other hand, down to $\mu_5\sim 100$ MeV, $\tilde\sigma$ width decreases and it becomes stable. The bumps seem to reflect the tachyonic nature of the decaying $\tilde\pi$ but they are not trivial to analyse. Finally $\tilde\eta'$ width grows up to the GeV scale, which shows that the effective theory is not applicable.

Secondly, we present the same result compared with a boosted decaying $\tilde\eta$. In the right panel of Fig. \ref{gamma_eta} one may observe that strong dependences on the 3-momentum appear but they have nothing to do with Lorentz time dilatation. It is remarkable that beyond $q=500$ MeV an extra bump arises at $\mu_5\approx 240$ MeV and grows rather fast. As to $\tilde\sigma$ and $\tilde\eta'$, no big differences occur when they are boosted.

\section{Vector Meson Dominance approach to LPB}

At this point, one may now wonder if such a mixing of states of different parities occurs in the vector/axial-vector sector. Many models are dealing with vector particles phenomenologically but if we assume that the vector mesons appear as part of a covariant derivative, no mixing term can be generated by operators of dimension 4 if $\mu_5$ is an isosinglet.
\medskip

Hence, in this work, vector mesons are introduced and treated in the conventional way using the Vector Meson Dominance model with no mixing of states with different parities. The only LPB effect will be the Chern-Simons term
\begin{equation}
\Delta\mathcal L\simeq \varepsilon^{\mu\nu\rho\sigma}\text{Tr}\left [\zeta_\mu V_\nu V_{\rho\sigma}\right ],
\nonumber
\end{equation}
where $\zeta_\mu\propto\mu_5\delta_{\mu 0}$. After implementing this extension, it may be found that vector mesons exhibit the following dispersion relation:
\begin{equation}
m_{V,\epsilon}^2-m_V^2\propto \epsilon\mu_5|\vec k|,
\end{equation}
where $\epsilon=0,\pm 1$ is the meson polarization and the breaking of Lorentz symmetry is explicit again. According to this result, massive vector mesons split into three polarizations with masses $m^2_{V,-} < m^2_{V,L}< m^2_{V,+}$, which is a clear signature of parity breaking. This effect may be investigated in the HIC phenomenology and in this work, we will focus in the $\rho^0\to e^+e^-$ decay channel.

\begin{figure}[h!]
\begin{minipage}[t]{9.2cm}
\centering
\hspace{-1em}\includegraphics[scale=0.35]{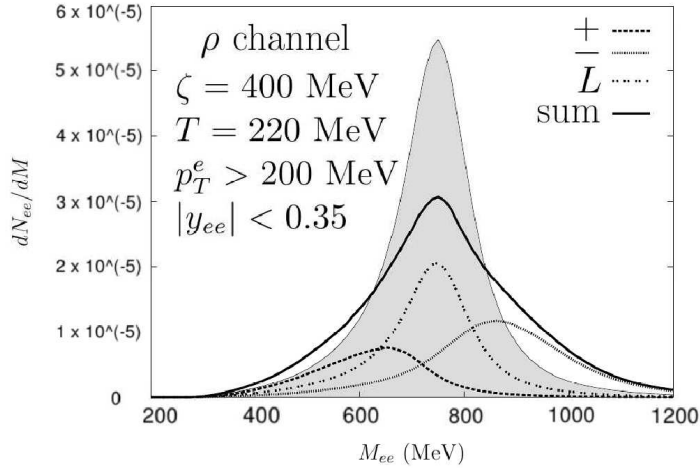}
\vspace{-1em}
\caption{QCD phase diagram depending on the temperature and density with the hypothetical region where LPB may occur.\label{pol_asymm}}
\end{minipage}\qquad \hspace{-0.3cm}\begin{minipage}[t]{5.45cm}
\vspace{-16em} \qquad In Fig. \ref{pol_asymm} we present a plot of the number of lepton pairs depending on their invariant mass. Therein we compare the QCD result preserving parity with our calculation for $\mu_5=290$ MeV. The separation of the different polarization contributions is displayed together with their sum in order to show the distorted shape of the parity breaking result. Moreover, it is remarkable the polarization asymmetry aside the peak and the fact that it could be experimentally measured.
\end{minipage}
\end{figure}

\vspace{-1.5em}

\section{Conclusions}
As we have seen, LPB is not forbidden by any physical principle in QCD at finite temperature and density. Therefore, we argued how topological fluctuations may be the origin of a parity breaking effect which is transmitted to hadronic physics via an axial chemical potential $\mu_5$. This phenomenon leads to unexpected modifications of the in-medium properties of both scalar and vector mesons. In the scalar sector, new eigenstates arise from the mixture of the known ones with interesting features such as the fact that $\tilde\pi$ and $\tilde\eta$ become tachyonic for big 3-momentum. Also $\tilde{\eta}$ acquires a non-negligible decay width meaning a significant indication for thermal equilibrium in the HIC fireball. Finally, in the vector meson sector, we found a distorted $\rho$ spectral function in comparison with the QCD expectation.

\end{document}